\documentclass[aps,twocolumn, floats,preprintnumbers,showpacs,prl]{revtex4}
\usepackage{graphicx}
\usepackage{url}
\usepackage{amsmath}
\usepackage{amsfonts}
\usepackage{amssymb}
\def\beq{\begin{equation}}
\def\eeq{\end{equation}}
\def\beqa{\begin{eqnarray}}
\def\eeqa{\end{eqnarray}}

\begin{document}
\preprint{FERMILAB-PUB-09-397-T}

\title{Dilaton-assisted Dark Matter}

\author{Yang Bai$^a$, Marcela Carena$^{a,b}$ 
and Joseph Lykken$^{a}$ 
\\
\vspace{2mm}
${}^{a}$Theoretical Physics Department, Fermilab, Batavia, Illinois 60510, \\
${}^{b}$Enrico Fermi Institute, Univ. of Chicago, 5640 Ellis Ave., Chicago, IL 60637
}


\begin{abstract}
A dilaton could be the dominant messenger between Standard Model fields and dark matter. The measured dark matter relic abundance relates the dark matter mass and spin to the conformal breaking scale. The dark matter-nucleon spin-independent cross section is
predicted in terms of the dilaton mass. We compute the current constraints on the dilaton from LEP and Tevatron experiments, and the gamma-ray signal from dark matter annihilation to dilatons that could be observed by Fermi-LAT.
\end{abstract}
\maketitle
{\it{\textbf{Introduction.}}}
Although there is compelling evidence for dark matter (DM) from astrophysical data, we still do not know the composition of dark matter or how dark matter particles interact with Standard Model (SM) particles.
Gravity is the default messenger, and indeed the existence of dark matter is inferred by its gravitational effects on ordinary matter and radiation.
It is possible that gravity is the unique messenger between the dark matter sector and the SM, as for example if the dark matter is a stable massive graviton of an extra dimensional theory~\cite{Feng:2003xh}; similar scenarios can accommodate gravitinos and other particles as ``superWIMP'' dark matter.
Another possibility is that dark matter particles are the lightest species of some hidden sector with its own
gauge interactions and some symmetry that stabilizes the dark matter. The coupling to SM fields is through
higher dimension operators suppressed by some heavy mediator scale, and the dark matter
particles can rather naturally be thermal relics with the observed relic density ~\cite{Feng:2008ya}.
Standard WIMP scenarios assume that particles of the electroweak sector, in particular the $Z$ boson, 
the Higgs, or particles of an extended Higgs sector, are the dominant messengers;
this is motivated by the dark matter candidates in models of supersymmetry~\cite{Jungman:1995df} and Universal Extra Dimensions~\cite{UED1}. 
A recently popular scenario is that the messenger is a new gauge boson, possibly light; the dark matter
particles are charged under this new gauge interaction, and at least some SM particles are as well,
either directly~\cite{Fox:2008kb} or through kinetic mixing~\cite{Pospelov:2007mp,ArkaniHamed:2008qn,Feldman:2007wj}.

Here we explore the possibility that a dilaton could be the dominant messenger particle. A dilaton can arise in many
scenarios for physics beyond the SM  as a pseudo-Goldstone boson from spontaneous breaking of scale invariance
in some sector of the full theory~\cite{Bardeen:1985sm,Goldberger:2007zk, Csaki:2007ns}. The dilaton acquires mass  from explicit
breaking of scale invariance in the otherwise scale-invariant sector, which we take to be small compared to the
energy scale $f$ of the spontaneous breaking. The dilaton will couple to the trace of the energy momentum tensor constructed
from all the fields in the scale-invariant sector, and may pick up additional couplings at loop level from the
scale anomaly.
As pointed out in~\cite{Goldberger:2007zk}, the simplest example of
such an effective dilaton is the Higgs boson of the SM, where the Higgs mass is small compared to $v= 246$ GeV, which in this case plays the role of $f$.

We will assume that the approximately scale-invariant sector includes both the dark matter particle and
some SM fields. If the dark matter particle has no direct couplings to SM fields we then expect the dilaton to be
the dominant messenger between the SM sector and the unknown dark matter sector. The dilaton field can easily be lighter than the dark matter particle and be the dominant decay product of dark matter annihilation. The dark matter thermal relic abundance, which is governed by its coupling to the dilaton, is thus strictly related to
the breaking scale $f$, for a fixed dark matter mass.

\begin{figure}[hb!]
\centerline{ \hspace*{0.0cm}
\includegraphics[width=0.45\textwidth]{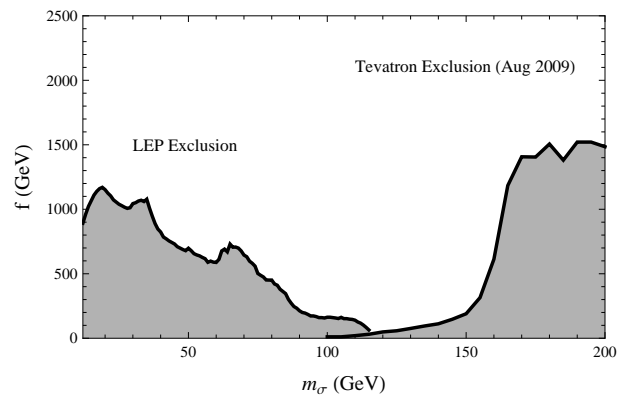}
}
\caption{The 95\% cl. constraints on the dilaton mass and the conformal breaking scale $f$ from direct searches at LEP and Tevatron.
\label{fig:exclusion}
}
\end{figure}

{\it{\textbf{Setup.}}}
We consider an effective theory with a cutoff $\sim4\pi f$, below which the scale symmetry is spontaneously broken.  
Let $\sigma$ denote the effective dilaton with mass $m_\sigma$ and use $\psi$ as a generic label for SM fermions. Then,
after electroweak symmetry breaking and to leading orders of powers of the dilaton field, we can write the
couplings of the dilaton to the SM fields~\cite{Goldberger:2007zk}
\beqa
{\cal L}_{\rm SM}&\supset&\frac{\sigma}{f}\left[ 2\,M^2_W\,W^{\mu+}\,W_{\mu-}+M_Z^2\,Z^\mu\,Z_\mu - \sum_{\psi}  m_{\psi}\,\bar{{\psi}}\,{\psi}     \right.    \nonumber \\
&& \hspace{-1.3cm} \left. -\frac{\alpha_{em}\,b_{em}}{8\,\pi}\,F_{\mu\nu}F^{\mu\nu} -\frac{\alpha_s\,b_{QCD}}{8\,\pi}\,G_{a\mu\nu}G^{a\mu\nu}
         \right]  -\frac{1}{2}\,m^2_\sigma \sigma^2   
         \label{eq:interactions1}
\eeqa
and to the DM matter field 
\beqa   
{\cal L}_{\rm DM}&\supset&  
\begin{cases}
 - \left(1+\frac{\sigma}{f} \right)\,m_\chi\,\bar{\chi}\,\chi & {\rm Fermion}  \vspace{2mm} \\ 
  -\left(1+\frac{2\sigma}{f} + \frac{\sigma^2}{f^2} \right)\frac{1}{2}\,m^2_\chi\, \chi^2  &   {\rm Scalar}  \vspace{2mm}\\
    \left(1+\frac{2\sigma}{f} + \frac{\sigma^2}{f^2} \right)\frac{1}{2}\,m^2_\chi\, \chi_\mu\chi^\mu  &   {\rm Gauge\; boson}\\
\end{cases}\,, 
\label{eq:interactions2}
\eeqa   
where $\chi$ is the dark matter field with various possible spins. Here, we have neglected derivative couplings of the dilaton to other fields, and for simplicity we only consider a single stable dark matter component .  For spin 1/2 dark matter we treat it as a Dirac particle, but a Majorana particle gives the same results in our later analysis. An obvious ${\cal Z}_2$ symmetry protects $\chi$ as a stable particle. 

The couplings of the dilaton to photons and gluons are generated at loop level by integrating out  particles heavier than the dilaton mass; these couplings should be proportional to contributions of the heavy particles to the beta functions. If the electromagnetic and QCD interactions are embedded in a conformal sector, the total beta functions vanish above the cutoff, so $b_{em}$ and $b_{\rm QCD}$ can be computed from the contributions of fermions lighter than the dilaton mass. Thus one has $b_{QCD}=-11+\frac{2}{3}\,n_f$ with $n_f$ as the number of quarks below $m_\sigma$, and $b_{em}=17/9$ when $m_W < m_\sigma < m_t$ and $11/3$ when $m_\sigma > m_t$. The other extreme is when QCD remains SM-like at the high scale and there are no additional heavy particles, so $b_{QCD}=-\frac{2}{3}$ when
$m_b < m_\sigma < m_t$. Our main results are computed assuming the ``conformal QCD" case, but we also comment on the
``SM-like QCD" case.

{\it{\textbf{Collider Constraints.}}}
The direct searches for the Higgs boson at LEP~\cite{Barate:2003sz} and the Tevatron~\cite{Tevatron:2009pt} can easily be adapted to constrain the dilaton, by making the substitution $v \to f$ in the electroweak and Yukawa couplings. In the conformal QCD scenario we also have to take into account the enhanced coupling of the dilaton field to gluons, by a factor $\simeq 11.5$ compared to the Higgs. As a result
the main dilaton decay channel is into two gluon jets for a dilaton mass below $2M_W$. The bounds on $f$ and the dilaton mass $m_\sigma$ are given in Fig.~\ref{fig:exclusion}, obtained by using the program {\tt Higgsbounds}~\cite{Higgsbounds}. 

At the Tevatron, the main sensitivity is in the dilepton channel from dilaton decays into two $W$'s. In this channel, the suppression of
the branching ratio is compensated by the enhancement of the gluon fusion production cross section. From Fig.~\ref{fig:exclusion}, one can see that there is still a widely allowed range for a light dilaton below 160~GeV even for $f$ not much above the electroweak scale. Other searches at Tevatron, such as the dijets resonance study~\cite{dijets}, do not impose significant additional constraints.

\begin{figure}[ht!]
\centerline{ \hspace*{0.0cm}
\includegraphics[width=0.48\textwidth]{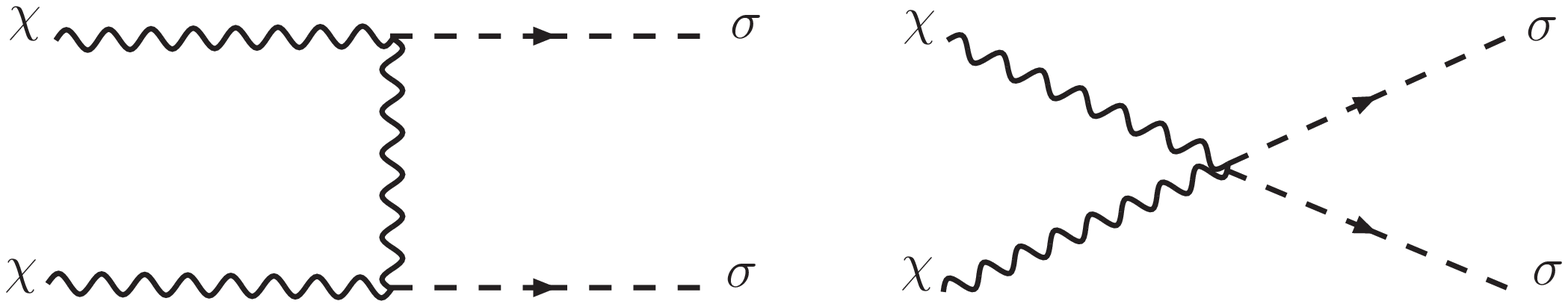}
}
\caption{The Feynman diagrams of the main dark matter annihilation channels for bosonic dark matter. The fermionic dark matter only has the left diagram.
\label{fig:Feynman}
}
\end{figure}

\begin{figure}[ht!]
\centerline{ \hspace*{0.0cm}
\includegraphics[width=0.45\textwidth]{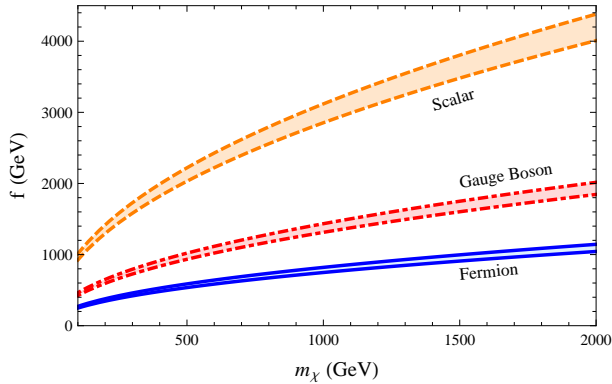}
}
\caption{The spontaneously conformal symmetry breaking scale as a function of the dark matter mass for different dark matter candidates. The dark matter relic abundance from WMAP is satisfied with $3\,\sigma$ bands. 
\label{fig:relic}
}
\end{figure}

{\it{\textbf{Relic Abundance.}}}
The dominant dark matter annihilation channels are into two dilaton particles, either via exchanging $\chi$ in the $t$-channel or directly using the operator containing two dilaton fields, as shown in Fig.~\ref{fig:Feynman}. 
We neglect diagrams with two SM particles in the final state and the diagrams using the dilaton cubic self-coupling, which are suppressed by $v/f$ and $m_\sigma/f$, respectively.  Following the standard calculation of the relic abundance, we expand the total annihilation rate in powers of $v$ as $\sigma v = a + b\,v^2 + \cdots$. Here, $v$, around $0.3\,c$ at the freeze-out temperature, is the relative velocity of the two annihilating dark matter particles. The updated value for the fraction of dark matter energy from WMAP is $\Omega_\chi\,h^2=0.1099\pm 0.0062$~\cite{WMAP}. The dark matter relic abundance is calculated as~\cite{Jungman:1995df}
\beqa
\Omega_\chi\,h^2\,\approx\,\frac{1.07\times 10^9}{{\rm GeV}\,M_{pl}\,\sqrt{g^*}}\frac{x_F}{a + 3(b -a/4)/x_F}\,,
\label{eq:relic}
\eeqa
with $x_F \equiv m_\chi/T_F$ and $T_F$ as the freeze-out temperature; $M_{pl}=1.22\times 10^{19}$~GeV is the Planck scale; $g^*$ is the number of degrees of freedom of relativistic particles at the freeze-out temperature and taken to be 86.25 if $T_F$ is around 50~GeV. 
\begin{table}[htdp]
\renewcommand{\arraystretch}{1.8}
\begin{center}
\begin{tabular*}{0.48\textwidth} {@{\extracolsep{\fill}} c  c  c c c   } 
\hline \hline
    & Fermion & Scalar  & Gauge Boson  \\ \hline
$a$  &  0 &    $\frac{9\,m_\chi^2}{16\,\pi\,f^4}$    &    $\frac{m_\chi^2}{36\,\pi\,f^4}$  \\  \hline
$b$   &     $\frac{3\,m_\chi^2}{128\,\pi\,f^4}$     &   $-\frac{25\,m_\chi^2}{128\,\pi\,f^4}$   &   $-\frac{19\,m_\chi^2}{864\,\pi\,f^4}$   \\
\hline \hline
\end{tabular*}
\end{center}
\caption{The two leading terms of the dark matter annihilation rate.}
\label{tab:ab}
\end{table}%
%
The freeze-out epoch satisfies the following iterative equation:
\beq
x_F\,=\,\ln{\left( \frac{5}{4}\sqrt{\frac{45}{8}}\frac{g}{2\pi^3}\frac{M_{pl}\,m_\chi(a + 6\,b/x_F)}{\sqrt{g^*}\sqrt{x_F}}             \right)} \,,
\label{eq:xf}
\eeq
with $g$ as the degrees of freedom of the dark matter particle. We list the formulas of $a$ and $b$ for different dark matter cases in Table~\ref{tab:ab} for $m_\chi \gg m_\sigma$, where the $p$-wave suppressions ($a=0$) for Majorana and Dirac fermionic dark matter are due to $CP$ and $P$ discrete symmetries, respectively.   
Using Eq.~(\ref{eq:relic}) and satisfying the experimental central value of $\Omega_\chi h^2$ from WMAP, we show the  symmetry breaking scale $f$ as a function of the dark matter mass in Fig.~\ref{fig:relic}.
In Fig.~\ref{fig:relic}, the solid and blue line, the dashed and orange line and the dot-dashed and red line are for fermion, scalar and gauge boson dark matter candidates, respectively. Here, the Dirac and Majorana fermions have approximately the same relation between $f$ and $m_\chi$. The only difference is their numbers of degrees of freedom, which only logarithmically change the freeze-out temperature. We can see from Fig.~{\ref{fig:relic}} that the dark matter mass is above the conformal symmetry breaking scale for part of those plots. This can be understood by a modestly large Yukawa coupling ${\cal O}(1\sim2)$ to the dark matter particle.

\begin{figure}[ht!]
\centerline{ \hspace*{0.0cm}
\includegraphics[width=0.45\textwidth]{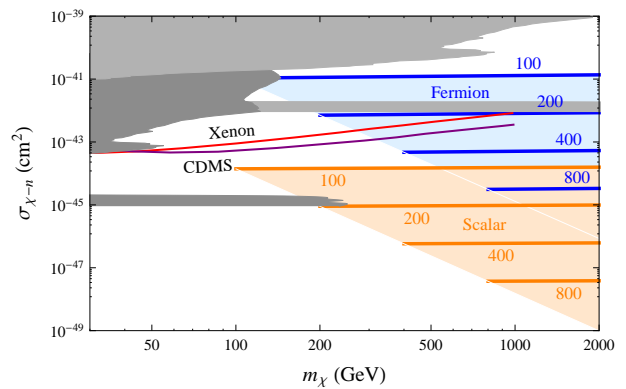}
}
\caption{Predicted spin-independent cross sections for the fermionic (blue and upper) and scalar (orange and lower) dark matter candidates. The gauge boson case is between the fermionic and scalar cases and is not shown here.  The red and purple lines are the latest exclusion limits from Xenon~\cite{Xenon} and CDMS~\cite{CDMS}. The gray and shaded region is excluded by direct dilaton searches at LEP and the Tevatron with the lighter one for fermionic dark matter and the darker one for scalar dark matter. The number with each solid line indicates the dilaton mass and in GeV. The dark matter mass is assumed to be larger than the dilaton mass. 
\label{fig:direct}
}
\end{figure}

{\it{\textbf{Direct Detection.}}}
Since the only detectable messenger between the dark sector and the SM sector is a scalar field, the dilaton, there are only spin-independent $\chi$-nucleon scattering processes.   
In the extreme non-relativistic limit and neglecting the nuclei form factor effects, we have the following formula for the DM-nuclei spin-independent elastic scattering cross section:
\beqa
\sigma_{\chi-N, SI}=\frac{s\,m_N^2 m_\chi^4}{\pi(m_\chi+m_N)^2 f^4 m_\sigma^4} [Z m_p +(A-Z)m_n]^2 \,.
\eeqa
Here, $s=4$ for fermionic dark matter and $s=1$ for bosonic dark matter; $m_N$, $Z$ and $A$ are the nuclei mass, charge and atomic number; $m_p$ and $m_n$ are the proton and neutron masses. Because the dilaton couples to the stress-energy tensor, its couplings to nucleons are proportional to their masses, and so are larger than the Higgs couplings to proton and neutron by approximately a factor of 3 in the conformal QCD scenario
Therefore in this scenario the dilaton as a messenger provides a factor of 9 increase for the direct detection cross section over a Higgs boson messenger for $f$ around the electroweak scale. For scattering an individual nucleon, we have the numerical $\chi$-nucleon cross section approximately as
\beqa
\sigma_{\chi-n, SI}&\approx& s\, \left(\frac{m_\chi}{500\, {\rm GeV}}\right)^2\,\left(\frac{1\, {\rm TeV}}{f}\right)^4\,\left(\frac{200\, {\rm GeV}}{m_\sigma}\right)^4 
\nonumber \\
&& \hspace{0cm}\,\times\,2\,\times\, 10^{-44}\, {\rm cm}^{2}\,.
\label{eq:direct}
\eeqa
Assuming that a single dark matter candidate saturates the observed relic abundance, we extract the required scale $f$ for each dark matter mass and spin from Fig.~\ref{fig:relic} and calculate the direct detection cross section. The resulting spin-independent $\chi$-nucleon scattering cross section is shown in Fig.~\ref{fig:direct}.
As can be seen from Fig.~\ref{fig:direct}, dark matter discovery prospects depend sensitively on the dilaton mass and on the spin of the dark matter particle. For a specific dilaton mass, the scattering cross section is almost independent of the dark matter mass.  This can be understood from the relic abundance formula in Eq.~(\ref{eq:relic}), which fixes $m_\chi^2/f^4$ because $x_F$ is around 25 for a wide range of dark matter masses. Therefore, the $\chi$-nucleon cross section only depends on the dilaton mass and the dark matter spin from Eq.~(\ref{eq:direct}). After satisfying the relic abundance, we have
\beqa
\sigma_{\chi-n, SI}&\approx& {\cal B}\,\left(\frac{200\, {\rm GeV}}{m_\sigma}\right)^4 
\,\times\,2\,\times\, 10^{-44} \,{\rm cm}^{2}\,,
\label{eq:direct2}
\eeqa
with ${\cal B}\approx (38, 1, 0.04)$ for fermion, gauge boson and scalar dark matter candidates.
If we assume SM-like QCD rather than the conformal QCD scenario, all of the cross sections
in Fig.~\ref{fig:direct} will be shifted downwards.

{\it{\textbf{Indirect Detection.}}}
Dark matter annihilation in our galaxy can produce energetic particles including photons, positrons, antiprotons and neutrinos. We concentrate on the diffuse gamma-ray predictions and leave other cosmic ray studies for future work. 
The differential gamma-ray flux from the dark matter annihilation is 
\beqa
\frac{d\Phi_\gamma}{dE_\gamma}\,=\,\frac{r_\odot\, \rho_\odot^2 \,\langle \sigma v_c\rangle}{4\,\pi\, m^2_\chi}\,\frac{d N_\gamma}{dE_\gamma}\,\bar{J}(\Delta \Omega)\Delta \Omega,
\label{eq:gammaray}
\eeqa
with $r_\odot = 8.5$~kpc and $\rho_\odot = 0.3$~GeV cm$^{-3}$ are the galactocentric distance of the solar system and the solar neighborhood DM density, $\langle \sigma v_c \rangle$ is the averaged dark matter annihilation rate in the current galaxy halo with $v_c \sim 10^{-3}$, and $\bar{J}$ measures the cuspiness of the galactic halo density profile. For the region around the galactic center, $\bar{J}$ can vary by a few orders of magnitude depending on the dark matter halo density profile. For illustration, we choose $\bar{J} \Delta \Omega \approx 0.7$, which corresponds to $\Delta \Omega = 2.4\times10^{-4}$~sr ($0.5^\circ \times 0.5^\circ$ about the galactic center) and the dark
matter profile of Navarro, Frenk and White (NFW)~\cite{NFW}. Since the dark matter velocity in the galactic halo is very small, the fermionic dark matter candidate, which has only $p$-wave annihilation, will produce a smaller gamma-ray flux than in the bosonic case by a factor of ${\cal O}(10^{-6})$.

\begin{figure}[ht!]
\centerline{ \hspace*{-0.1cm}
\includegraphics[width=0.45\textwidth]{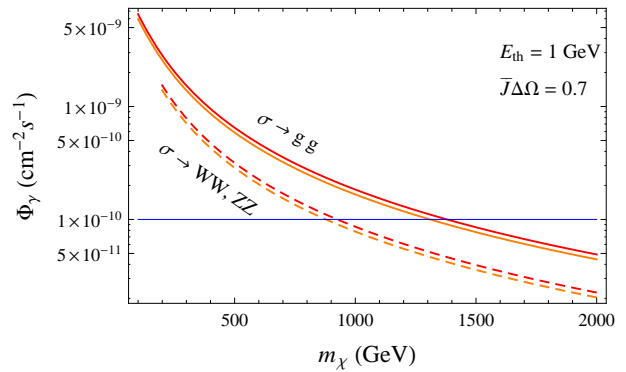}
}
\caption{The predicted photon integrated flux with a $1$~GeV threshold energy. The solid two lines are the predicted photon fluxes for the dilaton field mainly decays into gluons, while the dashed lines are for the $W, Z$ gauge bosons. The red (upper) and orange (lower) lines correspond to the gauge boson and the scalar dark matter candidates, respectively. The fermionic dark matter case is $p$-wave suppressed and not shown in this plot. The blue line is the potential discovery limit at Fermi LAT~\cite{FermiLAT}.
\label{fig:gammaray}
}
\end{figure}

The detailed photon flux also depends on the dilaton mass. If $2\,m_b < m_\sigma < 2\,M_W$, the dilaton dominantly decays to two gluons with a small branching ratio into two $b$'s. If $2\,M_Z < m_\sigma < 2\,m_t$, the dilaton mainly decays to $W, Z$ gauge bosons. We use $\tt PYTHIA$~\cite{Sjostrand:2006za} to simulate the gamma-ray fragmentation functions from the bottom quarks~\cite{Bai:2009ka}, from the gluons and from the $W, Z$ gauge boson in the dark matter annihilation final state. After substituting those numerically fitted functions into Eq.~(\ref{eq:gammaray}), we have the model predicted gamma-ray flux in Fig.~\ref{fig:gammaray}.
From Fig.~\ref{fig:gammaray}, we can see that the predicted photon integrated fluxes are approximately the same for the gauge boson (red and upper) and the scalar (orange and lower) dark matter candidates. 


{\it{\textbf{Conclusions.}}}
A striking feature of dilaton-assisted dark matter is that the dark matter direct detection rate is almost independent of the dark matter mass, and instead is determined by the dilaton mass and the dark matter spin. If a 100 GeV dilaton is discovered at the LHC, consistency with 
$\sigma_{\chi-n, SI}$ as measured in direct detection will confirm or disfavor this scenario. If both the dilaton mass and $\sigma_{\chi-n, SI}$ are measured precisely, we can even determine the spin of the dark matter candidate from Fig.~\ref{fig:direct}. If the scale $f$
can be accurately extracted from LHC data, there is also the possibility of determining the dark matter particle mass, even
when it is too heavy to extract from either direct detection or observe from LHC collisions.

\acknowledgments 
\vspace*{0.1in}
We thank Tom Appelquist, Bill Bardeen and Paddy Fox for interesting discussions. We also thank the Aspen Center for Physics where part of this work was finished.
Fermilab is operated by Fermi Research Alliance, LLC under contract no. DE-AC02-07CH11359 with the United States Department of Energy.


\vspace*{-.1in}
\begin{thebibliography}{99}
\vspace*{-.15in}

\bibitem{Feng:2003xh}
 J.~L.~Feng {\it et al.},
  Phys.\ Rev.\ Lett.\  {\bf 91}, 011302 (2003).

\bibitem{Feng:2008ya}
  J.~L.~Feng and J.~Kumar,
  Phys.\ Rev.\ Lett.\  {\bf 101}, 231301 (2008).

\bibitem{Jungman:1995df}
G.~Jungman {\it et al.},
  Phys.\ Rept.\  {\bf 267}, 195 (1996).
  
\bibitem{UED1}
  T.~Appelquist {\it et al.},
  Phys.\ Rev.\  D {\bf 64}, 035002 (2001).
    
  \bibitem{Fox:2008kb}
  P.~J.~Fox and E.~Poppitz,
  Phys.\ Rev.\  D {\bf 79}, 083528 (2009).
  
\bibitem{Pospelov:2007mp}
  M.~Pospelov {\it et. al.},
  Phys.\ Lett.\  B {\bf 662}, 53 (2008).
  
  \bibitem{ArkaniHamed:2008qn}
 N.~Arkani-Hamed {\it et al.},
  Phys.\ Rev.\  D {\bf 79}, 015014 (2009).
  
\bibitem{Feldman:2007wj}
  D.~Feldman {\it et al.},
  Phys.\ Rev.\  D {\bf 75}, 115001 (2007).
  
\bibitem{Bardeen:1985sm}
  W.~A.~Bardeen {\it et al.},
  Phys.\ Rev.\ Lett.\  {\bf 56}, 1230 (1986).
  
  \bibitem{Goldberger:2007zk}
  W.~D.~Goldberger {\it et al.},
  Phys.\ Rev.\ Lett.\  {\bf 100}, 111802 (2008).
  
\bibitem{Csaki:2007ns}
    C.~Csaki {\it et al.},
  Phys.\ Rev.\  D {\bf 76}, 125015 (2007).

  
\bibitem{Barate:2003sz}
  R.~Barate {\it et al.},
  Phys.\ Lett.\  B {\bf 565}, 61 (2003).
  
\bibitem{Tevatron:2009pt}
  CDF Note hwwmenn-090710 and D0 Note 5871-CONF.
  
\bibitem{Higgsbounds}
   P.~Bechtle {\it et al.},
  arXiv:0811.4169 [hep-ph].
  
\bibitem{dijets}
  T.~Aaltonen {\it et al.},
  Phys.\ Rev.\  D {\bf 79}, 112002 (2009).

\bibitem{WMAP}
  E.~Komatsu {\it et al.},
  Astrophys.\ J.\ Suppl.\  {\bf 180}, 330 (2009).
  
\bibitem{Xenon}
  J.~Angle {\it et al.},
  Phys.\ Rev.\ Lett.\  {\bf 100}, 021303 (2008).


\bibitem{CDMS}
  Z.~Ahmed {\it et al.},
  Phys.\ Rev.\ Lett.\  {\bf 102}, 011301 (2009).
  
\bibitem{NFW}
     J.~F.~Navarro {\it et al.},
  Astrophys.\ J.\  {\bf 490}, 493 (1997).
  
\bibitem{Sjostrand:2006za}
     T.~Sjostrand {\it et al.},
  JHEP {\bf 0605}, 026 (2006).
  
\bibitem{Bai:2009ka}
   Y.~Bai {\it et al.},
    Phys.\ Rev.\  D {\bf 80}, 055004 (2009).
  
  \bibitem{FermiLAT}
  \url{http://www-glast.stanford.edu/}
  
  
 



\end{thebibliography}
\end{document}